\documentclass{article}

\usepackage{stmaryrd}
\usepackage{amssymb,amsthm,amsmath,array}
\usepackage{appendix}
\usepackage{graphicx}
\usepackage[caption=false,font=footnotesize]{subfig}
\usepackage{xspace}
\usepackage[sort&compress, numbers]{natbib}
\usepackage{xcolor}
\usepackage{todonotes}
\usepackage{mathtools}
\usepackage{float}

\usepackage{standalone}
\graphicspath{{Figures/}}

\usepackage{tikz}
\usetikzlibrary{arrows}
\usepackage{pgfplots}
\pgfplotsset{compat=1.7}

\newcommand{\bs}{\boldsymbol}
\newcommand{\bb}{\mathbb}
\newcommand{\cl}{\mathcal}

\newcommand{\ts}{\textstyle}
\newcommand{\ie}{\emph{i.e.},\xspace}
\newcommand{\eg}{\emph{e.g.},\xspace}

\usepackage[top=1in, left=1in, right=1in, bottom=1in]{geometry}
\renewenvironment{abstract}{\bf {\em\ Abstract---}}{}

\makeatletter
\renewcommand{\section}{\@startsection {section}{1}{\z@}%
             {-3.5ex \@plus -1ex \@minus -.2ex}%
             {2.3ex \@plus.2ex}%
             {\normalfont\large\bfseries}}
\makeatother
\begin{document}
\title{Compressive Scanning Transmission Electron Microscopy}
%
\author{
	D.~Nicholls\textsuperscript{1}, 
	A.~Robinson\textsuperscript{1}, 
	J.~Wells\textsuperscript{2},
	A.~Moshtaghpour\textsuperscript{1, 3},
	M.~Bahri\textsuperscript{1},\\
	A.~Kirkland\textsuperscript{3,6}, and N.~Browning\textsuperscript{1,4,5}\footnote{
		DN, AR, JW, and NB are funded by EPSRC. JW is also funded by Sivananthan Laboratories. AM is funded by the RFI.
	    }\\ 
	\fontsize{9}{1}\selectfont \textsuperscript{1} Department of Mechanical, Materials and Aerospace Engineering, University of Liverpool, UK.\\
	\fontsize{9}{1}\selectfont \textsuperscript{2} Distributed Algorithms Centre for Doctoral Training, University of Liverpool, UK.\\
	\fontsize{9}{1}\selectfont \textsuperscript{3} Correlated Imaging group, Rosalind Franklin Institute, Harwell Science and Innovation Campus, Didcot, UK.\\
	\fontsize{9}{1}\selectfont \textsuperscript{4} Physical and Computational Science Directorate, Pacific Northwest National Laboratory, Richland, USA.\\
	\fontsize{9}{1}\selectfont \textsuperscript{5} Sivananthan Laboratories, 590 Territorial Drive, Bolingbrook, IL, USA.\\
	\fontsize{9}{1}\selectfont \textsuperscript{6} Department of Materials, University of Oxford, UK.
	}\date{\empty}
%
\maketitle
\begin{abstract}
Scanning Transmission Electron Microscopy (STEM) offers high-resolution images that are used to quantify the nanoscale atomic structure and composition of materials and biological specimens. In many cases, however, the resolution is limited by the electron beam damage, since in traditional STEM, a focused electron beam scans every location of the sample in a raster fashion. In this paper, we propose a scanning method based on the theory of Compressive Sensing (CS) and subsampling the electron probe locations using a line hop sampling scheme that significantly reduces the electron beam damage. We experimentally validate the feasibility of the proposed method by acquiring real CS-STEM data, and recovering images using a Bayesian dictionary learning approach. We support the proposed method by applying a series of masks to fully-sampled STEM data to simulate the expectation of real CS-STEM.
Finally, we perform the real data experimental series using a constrained-dose budget to limit the impact of electron dose upon the results, by ensuring that the total electron count remains constant for each image. 
\end{abstract}

\vspace{2mm}

\section{Introduction}
\label{sec:intro}

\begin{figure}[t]
    \centering
    \scalebox{0.53}{\includegraphics{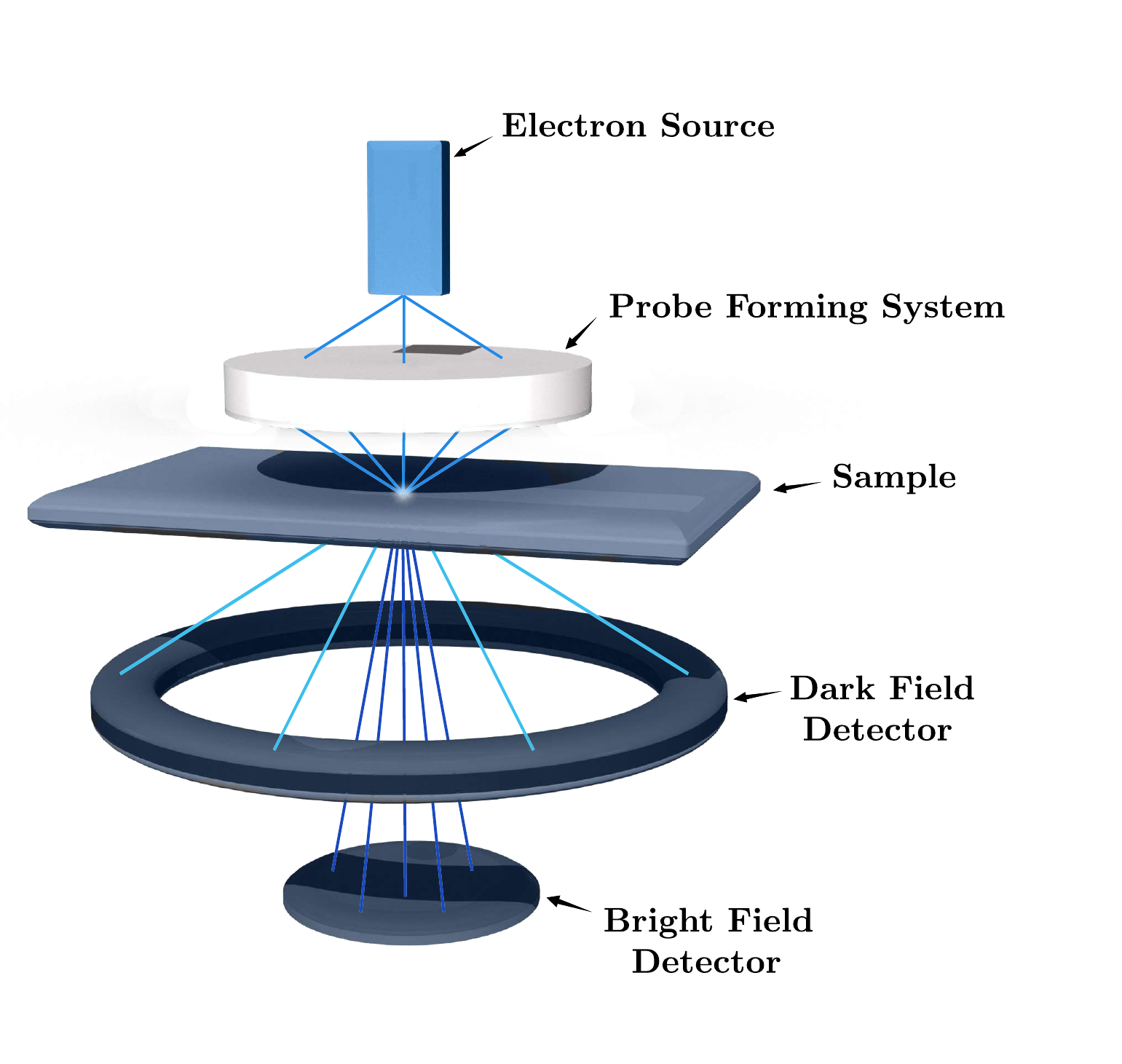}}
    \caption{Operating principles of STEM. Electrons are emitted from the source and subsequently condensed by the probe forming system. The probe is then raster scanned over the desired sampling area and the transmitted electrons are detected by either a dark field or bright field detector. The formed image is the intensity of the resulting electron wave-function at each pixel, corresponding to a certain set of scan coordinates.}
    \label{fig:stem-scheme}
\end{figure}

Advances in Scanning Transmission Electron Microscopy (STEM) over the last 30 years have come at a cost. With the advent of evermore efficient and brighter electron guns, aberration correctors condensing the electron probe down to the angstrom level, and the desire to image single atoms with precision and accuracy, electron beam probes have become incredibly intense. As schemed in Fig.~\ref{fig:stem-scheme}, a STEM consists of  an electron source, a probe forming system, a scan coil system, and an image forming system. STEMs operate on one basic principle; an electron probe (with typical size of less than 0.1nm) is rastered over a material, and the resultant electron-sample interaction is measured at each location in the sample. This measurement can take many forms; imaging via the collection of the transmitted or scattered electrons, or spectroscopy by the collection of x-rays (EDS), or measuring the energy loss in the electron beam (EELS). 

The intense probes used, while performing admirably for robust metals and crystals, have imposed a new limitation that is becoming more and more commonplace in the field of electron microscopy characterisation - beam damage. Rather than being limited by the resolution of the instrument, analysis is now being limited by the stability of the sample and its ability to resist change under the power of the electron beam.

Electron beam damage has been well studied \cite{egerton2021dose}, and is especially well understood by those wishing to image biological materials \cite{peet2019energy, wu2020liquid}, wherein cryo-electron microscopy is necessary to image cells and tissues which are very beam sensitive. Advances are also being made with electron detectors to increase their sensitivity, such that fewer electrons are needed to form images. These methods are often very specialised and require a high degree of investment in terms of equipment and training, and often still struggle with poor signal-to-noise ratios. 

An alternative without compromising the resolution is through \textit{subsampling}. Subsampling is performed in STEM by manipulating the regular scanning coils to move the electron probe in such a way as to forming incomplete scans. This incomplete signal is then recovered using an appropriate reconstruction method that depends on the signal type being acquired. Not only does subsampling reduce the electron-dose needed to acquire images, it also increases the rate of image acquisition. This is especially useful for those interested in time-dependent phenomena and \textit{in-situ} characterisation \cite{mehdi2019controlling}. There is also the added benefit of reducing the effects of time-related instabilities, such as sample drift, and as a result may actually increase the quality of measurements over traditional means.

The theory of Compressive Sensing (CS) \cite{donoho2006compressed,candes2006robust} applied to STEM operation has already shown significant capability to image samples that are traditionally unattainable. The reduction of electron dose due to the reduced STEM measurements allowed High-Angle Annular Dark Field (HAADF) images of calcium carbonate (CaCO$_{3}$) to be taken without significant amorphization of the sample \cite{kovarik2016implementing}. Furthermore, CS has been applied to multi-dimensional STEM techniques such as 3D tomography \cite{genc2013xeds,donati2017compressed} and ptychography \cite{humphry2012ptychographic} to allow not only high resolution imaging of sensitive samples, but also to increase the rate at which data can be acquired. In ptychography, subsampling can be applied to both the real and reciprocal spaces; increasing the speed of acquisition \cite{stevens2018subsampled}.

In this paper we resort to the theory of Compressed Sensing (CS) to reduce the electron beam damage in STEM while preserving the quality of the reconstructed images. Our idea is to subsample the electron probe location along its trajectory, hence reducing the electron dose proportional to the subsampling ratio. We highlight the limiting factors that does not allow every subsampling strategy to be practically feasible. The proposed CS-STEM follows line hop sampling scheme: subsampling (at random) the adjacent locations to the probe's (by default) line trajectory. See Fig.~\ref{fig:results_constrained} for some examples. The image recovery problem is turned into a Bayesian dictionary learning problem based on the Beta Process Factor Analysis developed in \cite{paisley2009nonparametric,zhou2009non}. As an advantage of such method, we do not need a priori knowledge about the noise variance and the dictionary in which the image has sparse/compressible representation. To speed up the inference process, we leverage a stochastic expectation maximisation \cite{sertoglu2015scalable} approach to infer the unknown parameters and in turn recover the image. We employ a constrained-dose budget for each of the subsampled images, such that the total electron count remains constant throughout. This way, the only data compression is in the spatial domain, and not the total number of electrons that are incident upon our sample. 

\section{Main Results: Compressive STEM}
\label{sec:cs-stem}
We now explain our CS-STEM framework that collects STEM measurements by subsampling the electron probe positions and reconstructs the sample image from those measurements by solving a Bayesian dictionary learning problem.

\subsection{Acquisition model}
Let $\bs x \in \bb R^N$ be the discretized and vectorized image (with $N$ pixels) of the sample located in the field of view. When operating in the raster mode\footnote{Alternative modes are the snake and Hilbert scans, whose sensing model reads \eqref{eq:raster-stem-sensing-model}, too \cite{velazco2020evaluation}.}, the electron probe scans every entry of $\bs x$, as shown in Fig.~\ref{fig:stem-scheme}, and the resulting scattered electrons are collected by the detector. Hence, in a simplified STEM model, the raster scan observations reads
\begin{equation}\label{eq:raster-stem-sensing-model}
    \bs y^{\rm raster} \coloneqq \bs x + \bs n^{\rm raster} \in \bb R^{N},
\end{equation}
where $\bs n^{\rm raster}$ models a noise. 

As mentioned in Sec.~\ref{sec:intro}, our CS-STEM system consists in acquiring fewer measurements to reduce the electron dose imposed on the sample. This can be achieved by subsampling $M < N$ probe positions indexed in a subset $\Omega \subset \{1,\cdots,N\}$ with $|\Omega| =M$; resulting in the following sensing model, \ie 
\begin{equation}\label{eq:cs-stem-sensing-model}
    \bs y^{\rm cs} \coloneqq \bs P_\Omega\bs x + \bs n \in \bb R^{N},
\end{equation}
where $\bs P_\Omega \in \{0,1\}^{N\times N}$ is a mask operator with $(\bs P_\Omega \bs x)_j = x_j$ if $j \in \Omega$, and  $(\bs P_\Omega \bs x)_j = 0$ otherwise. 

Alongside the number of sampled probe positions $M$, sampling pattern or trajectory is another constraint in practical STEM. For example, moving the electron probe over the large distances yields poor quality images due to scan coil hysteresis (the inability of the electromagnets to return to their state prior to inducing the initial field). Therefore, subsampling the probe positions purely at random, \ie desired by CS theory, using regular scan coils becomes unpleasant. A solution to reduce the hysteresis problem would be the use of a fast beam blanker along with a suitable scan generator, however, at the expense of increased scan time. Random line scanning can overcome the hysteresis issue but at the cost of increased accumulation of electron dose in a pixel (beam overlap).

Our CS-STEM operates using line hop sampling: random variation perpendicular to the scan direction, which has shown successful result in reducing both scan coil hysteresis and beam overlap \cite{nicholls2020minimising,nicholls2021subsampled}.

\subsection{Image recovery method}
Like any other CS application, recovering an image from subsampled measurements in  \eqref{eq:cs-stem-sensing-model} requires a suitable low-complexity prior model of the images. In this work, we assume that the STEM images are sparse/compressible in a dictionary. For the sake of universality and instead of using an off-the-shelf dictionaries (\eg wavelets)  we further assume that the dictionary is unknown and should be estimated for each experiment; hence, amounting to a blind dictionary learning approach for our inpainting problem. 

In this paper we perform dictionary learning adopting a Bayesian non-parametric method called Beta Process Factor Analysis (BPFA) and introduced in \cite{paisley2009nonparametric}. As an advantage of such non-parametric approach, we do not need to assume a priori knowledge about the noise variance (unlike K-SVD approach~\cite{elad2006image}) and sparsity level of the signal in the dictionary. We cover here only the required elements of BPFA and refer the reader to \cite{paisley2009nonparametric,zhou2009non} for a detailed survey. See also \cite{huang2014bayesian} for an application of BPFA in the context of compressive MRI.

Given a CS-STEM measurement $\bs y^{\rm cs}$, we first partition it into $N_p$ overlapping patches $\{\bs y_i\}_{i=1}^{N_p}$, with each patch $\bs y_i \in \bb R^{B^2}$; hence, resulting in $\ts N_p = (\sqrt{N}-B+1)^2$ total number of patches. Similarly, we partition the sample image, mask operator, and noise as $\{\bs x_i\}_{i=1}^{N_p}$, $\{\bs P_{\Omega_i}\}_{i=1}^{N_p}$, and $\{\bs n_i\}_{i=1}^{N_p}$ respectively, such that for each patch $i \in \{1,\cdots,N_p\}$,
\begin{equation} \label{eq:cs-stem-patch-sensing-model}
    \bs y_i = \bs P_{\Omega_i}\bs x_i + \bs n_i \in \bb R^{B^2}.
\end{equation}

Furthermore, we assume that each image patch is sparse in a shared dictionary, \ie $\bs x_i = \bs D \bs \alpha_i$, where $\bs D \in \bb R^{B^2\times K}$ denotes the dictionary with $K$ atoms and $\bs \alpha_i \in \bb R^{K}$ is a sparse vector of weights or coefficients. Unlike traditional sparse coding approaches, which require a pre-defined dictionary or at least the number of dictionary atoms, we here desire to jointly learn the shared dictionary and weights, given the CS-STEM measurements. To achieve that goal we a BPFA approach that allows us to infer $\bs D$, $\bs \alpha$, $K$, and the noise statistics and in turn reconstruct the sample image.

BPFA assumes that \textit{(i)} the dictionary atoms $\{\bs d_k\}_{k=1}^{K}$ are drawn from a zero-mean multivariate Gaussian distribution; \textit{(ii)} both the components of the noise vectors $\bs n_i$ and the non-zero components of the weight vectors $\bs \alpha_i$ are drawn \textit{i.i.d.} from zero-mean Gaussian distributions; \textit{(iii)} the sparsity prior on the weights is promoted by the Beta-Bernoulli process~\cite{paisley2009nonparametric}. Mathematically, for all $i \in \{1,\cdots,N_p\}$ and $k \in \{1,\cdots,K\}$,
\begin{subequations}
\begin{align}
    \bs y_i &= \bs P_{\Omega_i} \bs D \bs \alpha_i + \bs n_i, \!& \bs \alpha_i &= \bs z_i \circ \bs w_i \in \bb R^K,\label{eq:bpfa-1}\\
    \ts \bs D &= [\bs d_1^\top, \cdots, \bs d_{K}^\top]^{\top}, \!& \bs d_k & \sim  \cl N(0, B^{-2} \bs I_{B^2}),\label{eq:bpfa-2}\\
    \bs w_i & \sim  \cl N(0, \gamma_w^{-1} \bs I_{K}), \!& \bs n_i & \sim \cl N(0, \gamma_n^{-1} \bs I_{B^2}),\label{eq:bpfa-3}\\
    \bs z_i &\sim \!\ts \prod_{k=1}^{K} {\rm Bernoulli}(\pi_k),\! & \pi_k &\sim\! {\rm Beta}(\ts \frac{a}{K}, \frac{b(K-1)}{K}),\label{eq:bpfa-4}
\end{align}
\end{subequations}
where $\bs I_K$ is the identity matrix of dimension $K$, $\circ$ denotes the Hadamard product, and $a$ and $b$ are the parameters of the Beta process. The binary vector $\bs z_i$ in \eqref{eq:bpfa-4} determines which dictionary atoms to be used to represent $\bs y_i$ or $\bs x_i$; and $\pi_k$ is the probability of using a dictionary atom $\bs d_k$. In \eqref{eq:bpfa-3}, $\gamma_w$ and $\gamma_n$ are the (to-be-inferred) precision or inverse variance parameters. It is common to place a non-informative, \ie flat, gamma hyper-priors on $\gamma_w$ and $\gamma_n$, by fixing them to small values \cite{tipping2001sparse}. The sparsity level of the weight vectors, \ie $\{\|\bs \alpha_i\|_0\}_{i=1}^{N_p}$ is controlled by the parameters $a$ and $b$ in \eqref{eq:bpfa-4}. However, as discussed in \cite{zhou2009non}, those parameters tend to be non-informative and the sparsity level of the weight vectors is inferred by the data itself. 

Unknown parameters in the model above can be inferred using Gibbs sampling \cite{zhou2009non}, variational inference \cite{paisley2009nonparametric}, or (as in this paper) Expectation Maximisation (EM) \cite{dempster1977maximum,sertoglu2015scalable}. In short, EM involves an expectation step to form an estimation of the latent variables, \ie $\{\|\bs \alpha_i\|_0\}_{i=1}^{N_p}$, and a maximisation step to perform a maximum likelihood estimation to update other parameters. Since the number of patches $N_p$ may be large, a stochastic (or mini-batch) EM approach is implemented, where the $N_p$ patches are (randomly) partitioned into batches of size $N_b$ and those batches are processed sequentially. Similar ideas have been used in \cite{sertoglu2015scalable,paisley2014bayesian}.

\section{Experiments}
\label{sec:sim}
\begin{figure}[t]
    \centering
    \scalebox{0.95}{\includegraphics{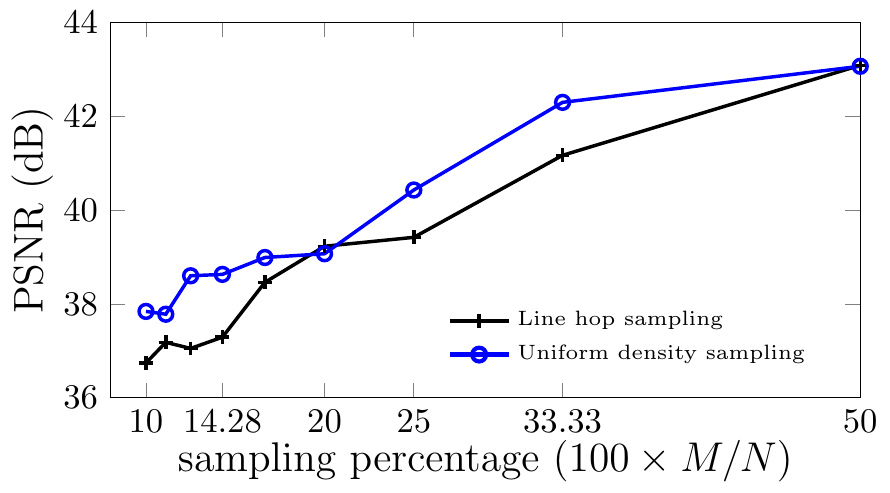}}
    \caption{Performance comparison between line hop sampling (desired in practice) and UDS (desired by the CS theory).}
    \label{fig:snr_curve}
\end{figure}

\begin{figure*}[t]
    \centering
    \begin{minipage}{\textwidth}
    \centering
        \begin{minipage}{0.19\textwidth}
        \centering
        \scalebox{0.92}{\includegraphics{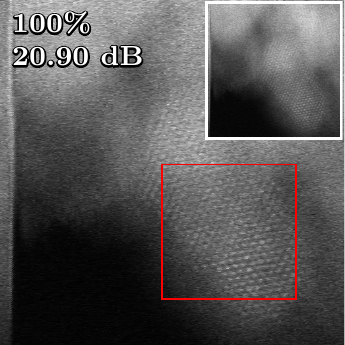}}
        \end{minipage}
        \begin{minipage}{0.19\textwidth}
        \centering
        \scalebox{0.92}{\includegraphics{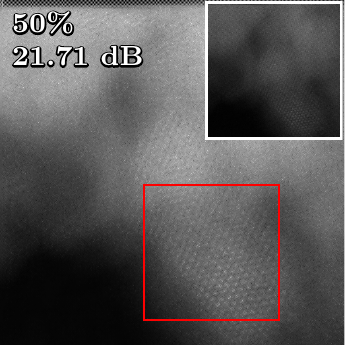}}
        \end{minipage}
        \begin{minipage}{0.19\textwidth}
        \centering
        \scalebox{0.92}{\includegraphics{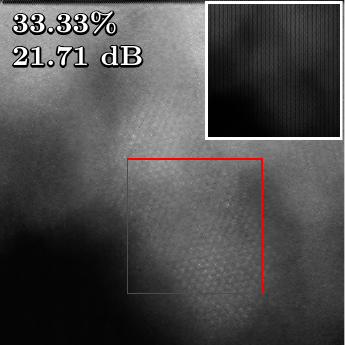}}
        \end{minipage}
        \begin{minipage}{0.19\textwidth}
        \centering
        \scalebox{0.92}{\includegraphics{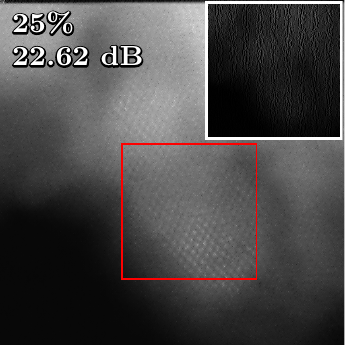}}
        \end{minipage}
        \begin{minipage}{0.19\textwidth}
        \centering
        \scalebox{0.92}{\includegraphics{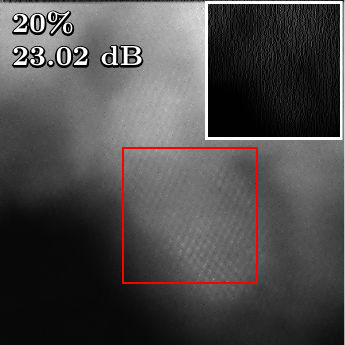}}
        \end{minipage}
    \end{minipage}
    \\
    \begin{minipage}{\textwidth} 
    \centering
        \begin{minipage}{0.19\textwidth}
        \centering
        \scalebox{0.92}{\includegraphics{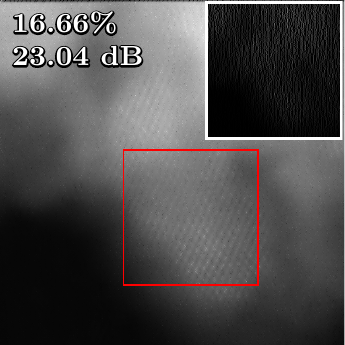}}
        \end{minipage}
        \begin{minipage}{0.19\textwidth}
        \centering
        \scalebox{0.92}{\includegraphics{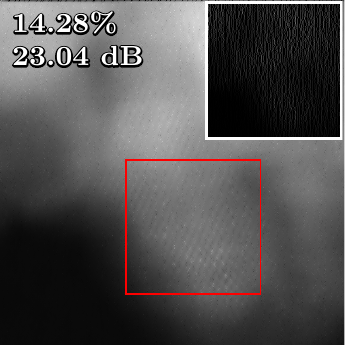}}
        \end{minipage}
        \begin{minipage}{0.19\textwidth}
        \centering
        \scalebox{0.92}{\includegraphics{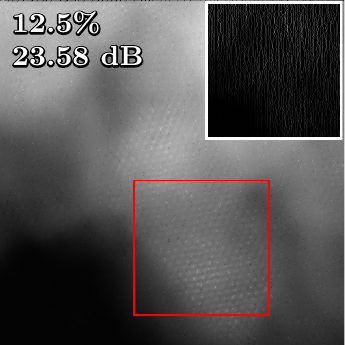}}
        \end{minipage}
        \begin{minipage}{0.19\textwidth}
        \centering
        \scalebox{0.92}{\includegraphics{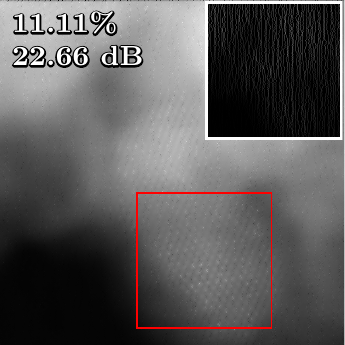}}
        \end{minipage}
        \begin{minipage}{0.19\textwidth}
        \centering
        \scalebox{0.92}{\includegraphics{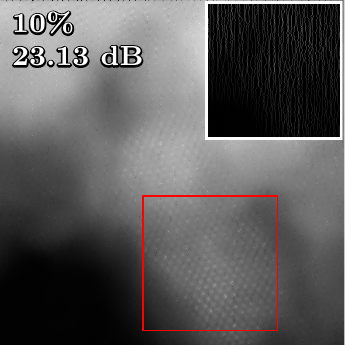}}
        \end{minipage}
    \end{minipage}
    \caption{Reconstruction quality of CS-STEM with \textit{constrained-dose budget}. Observations are overlaid on top-right of each reconstructed image. PSNR values, \ie reported on top-right of each image, are computed between the reference crop (from STEM observations obtained with $t_d = 40\mu$s, not shown here) and the crops identified by red squares. The electron dose budget is constrained to $4\mu s \cdot N$; thus, for each sampling ratio the dwell time is adjusted such that $t_{d}\cdot M = 4\mu s \cdot N$. We observe that sub-sampling the electron probe with higher dwell time improves the image quality.}.
    \label{fig:results_constrained}
\end{figure*}

We conducted several experiments to investigate the impact of sampling ratio on the quality of image reconstruction. The experiments were carried out using a JEOL 2100F aberration corrected STEM with a beam voltage of 200 kV. Throughout the experimental series, alignment was maintained, and focus conditions were kept similar to reduce inconsistency in measurements. The sample of interest is heat treated ceria (CeO$_{2}$), and the same nanoparticle of this composition was kept in view throughout the experimental series.

In the first set of experiments we compare line hop sampling with Uniform Density Sampling (UDS). A full raster scan was performed with a dwell time of $t_d = 40 \mu$s. A patch of $256\times 256$ pixels was denoised via BPFA\footnote{We initialized the dictionary elements using a normal distribution and set $K=512$ and $B = 30$ without performing rigorous parameter tuning.} algorithm and the resulting image was considered as the ground truth. We then simulated the CS-STEM measurements by applying artificial subsampling masks, \ie generated following line hop sampling and UDS schemes, on the ground truth image. This procedure is repeated over
10 random realisations of the subsampling mask. The averaged reconstruction quality measured as Peak Signal-to-Noise-Ratio (PSNR) for different sampling ratios $M/N$ is plotted in Fig.~\ref{fig:snr_curve}. We observe that the (theoretically-approved) UDS scheme outperforms the (practically-feasible) line hop sampling scheme. Despite observing approximately 1 dB PSNR gap, line hop sampling still provides high quality results, \eg achieves PSNR = 36.74 dB for $M/N = 10\%$.

In the second set of experiments we intend to prove the concept of CS-STEM with constrained-dose budget. It can be shown that for a constant electron current, the total electron count $N_e \propto t_{d} \cdot M$. We refer readers to~\cite{egerton2021dose} for more information regarding electron dose rates in STEM operation.  In this context, we first acquired a reference STEM image with $N = 512^2$ pixels and dwell time $t_d = 4 \mu$s. In order to ensure that a constant electron count was used for each sampling percentage, we varied the dwell time according to the sampling ratio $(M/N)$. The CS-STEM observations were further taken at sampling ratios of $M/N = \{50,\cdots,10\}\%$ and dwell times of $t_d = \{8,\cdots,40\}\mu$s, respectively. Images were then reconstructed by BPFA. Throughout the experiment, no structural changes to the sample were observed. However, as evident in Fig.~\ref{fig:results_constrained}, sample drifting from one acquisition to another was inevitable. To evaluate the reconstruction quality, we define the reference image as a $200 \times 200$-pixel crop of the STEM observations acquired with $t_d = 40\mu$s, which is the highest quality observation. To account for the sample drift, we computed the PSNR between that reference patch and every $200\times 200$-pixel patch of the reconstructed image. The patch that yields the highest PSNR value is considered as the matched patch for the reference patch and the corresponding PSNR value is reported.

Fig.~\ref{fig:results_constrained} illustrates the observations and their corresponding reconstructed image. These results stress that for a fixed electron-dose budget an application of CS-STEM results in a higher quality of reconstructed images. Despite the smoothed images for very low sampling ratios, \eg $M/N = 10\%$, we observe that the atomic structure of the sample, which is of high importance for microscopists, is successfully preserved.

\section{Conclusion}
\label{sec:conclusion}
In conclusion, we have demonstrated a practical application of CS in STEM, and the ability for dictionary learning methods such as BPFA to reconstruct high quality images from real CS-STEM measurements obtained with line hop sampling and with a constrained-dose budget. Furthermore, we have shown that not only could this provide an invaluable tool for imaging beam-sensitive materials and time-dependent phenomena as well as reducing acquisition time, but that the development of other sampling regimes could theoretically improve results at even lower electron doses. These methods may be applied to a wide variety of data, such as EDS maps, EELS spectra, and multidimensional datacubes such as 3D tomography and hyperspectral data.
Future work to improve the practical application of CS-STEM will likely require a reduction of inference time, \eg using deep learning methods allowing for real-time reconstructions, and thus enabling the user to make any required microscope adjustments with only an initially subsampled measurement. 

\vfill\pagebreak
\bibliographystyle{IEEEtran}        
\bibliography{refs}

\end{document}